\documentclass[a4paper,11pt]{article}
\pdfoutput=1 

\usepackage{booktabs}
\usepackage{multirow}
\usepackage{jinstpub} 

\title{\boldmath The LSPE-Strip beams}

\author[a,b,1]{S. Realini,\note{Corresponding author.}}
\author[a,b]{C. Franceschet,}
\author[c]{F. Villa,}
\author[c]{M. Sandri,}
\author[d]{G.Addamo,}
\author[e, f]{P. Alonso-Arias,}
\author[a, b]{M. Bersanelli,}
\author[c]{F. Cuttaia,}
\author[g]{M. Jones,}
\author[h]{M. Maris,}
\author[k]{F. P. Mena,}
\author[a, b]{A. Mennella,}
\author[l]{R. Molina,}
\author[c]{G. Morgante,}
\author[a, b]{M. Tomasi,}
\author[i, j]{M. Zannoni}

\affiliation[a]{Universit\`a degli Studi di Milano, via Celoria 16, Milano, Italy}
\affiliation[b]{INFN–Sezione di Milano, Via Celoria 16, Milano, Italy}
\affiliation[c]{INAF–OAS Bologna, Via P. Gobetti 101, Bologna, Italy}
\affiliation[d]{IEIIT-CNR, Politecnico di Torino, Corso Duca degli Abruzzi 24, Torino, Italy}

\affiliation[e]{Instituto de Astrofísica de Canarias, Calle Vía Láctea SN, La Laguna, Spain}
\affiliation[f]{Departamento de Astrofísica, Universidad de La Laguna, La Laguna, Spain}

\affiliation[g]{University of Oxford, Denys Wilkinson Building, Keble Road, Oxford OX1 3RH, UK}

\affiliation[h]{INAF–Osservatorio Astronomico di Trieste, Via G.B. Tiepolo 11, Trieste, Italy}

\affiliation[i]{Universit\`a degli Studi di Milano-Bicocca, Piazza della Scienza 3, 20126 Milano, Italy}
\affiliation[j]{INFN–Sezione di Milano Bicocca, Piazza della Scienza 3, 20126 Milano, Italy}
\affiliation[k]{Electrical Engineering Department, University of Chile, Tupper 2007, Santiago, Chile}
\affiliation[l]{Astronomy Department, University of Chile, Camino El Observatorio 1515, Santiago, Chile}

\emailAdd{sabrina.realini@unimi.it}

\abstract{In this paper we describe the design and characterization of the optical system of LSPE/Strip, a coherent polarimeter array that will observe the microwave sky from the Teide Observatory in Tenerife in two frequency bands centred at 43 and 95 GHz through a dual-reflector crossed-Dragone telescope of 1.5 m aperture. 
In general, optical systems composed by a telescope-feed array assembly have non-idealities that might limit their ability to perform high-precision measurements. It is thus necessary to understand, characterize and properly control these systematic effects.
For this reason, we performed electromagnetic simulations to characterize angular resolution, sidelobes, main beam symmetry, polarization purity and feedhorns orientation.
The results presented in this paper will be an essential input for further optical studies and for the LSPE/Strip data analysis. Ultimately, they will be used to assess the impact of optical systematic effects on the scientific results.}


\proceeding{The LSPE/Strip instrument description and testing}

\begin{document}
\maketitle
\flushbottom

\section{Introduction}
\label{sec:intro}
The Strip instrument is part of the Large Scale Polarization Explorer, which is one of the upcoming experiments devoted to the observation of the Cosmic Microwave Background (CMB) polarization on large angular scales (see \cite{Piacentini2020}). The main goal of Strip is the characterization of the Galactic synchrotron emission in the Q-band.
Strip is based on a coherent polarimeter array made of 49 receivers in the Q-band (centred at 43 GHz) and 6 receivers in the W-band (centred at 95 GHz). The latter array will be used to monitor and study the atmospheric emission at the observing site. These receivers are coupled with a 1.5 m off-axis dual reflector telescope by an array of corrugated feedhorns \cite{Franceschet2021}. 
The optics is composed by a telescope-feed array assembly designed to achieve an angular resolution of $\sim 20$ arcmin in Q band with very low sidelobes. While the optical system is optimized, approximate beam knowledge or unwanted straylight due to the sidelobes pick-up may introduce residual systematic effects in the measurements. 

We present the electromagnetic model of the Strip telescope, which includes all the elements that affect the overall optical response. To obtain an accurate estimate of the effective beam in the instrument observing frequency band, main beam parameters have been calculated at several frequencies within the $\sim 20$\% nominal bandwidth. Furthermore, we have analysed the impact on beam properties of surface  distortions in the primary and secondary mirrors. 
Main beams simulations are based on reflections and diffractions from the primary and secondary mirrors, while we carried out a separate analysis for the sidelobes region to include also the baffles.

Accurate predictions and measurements of the beam shape are essential both during the instrument development phase to optimize the design of each feedhorn, as well as their location and orientation in the focal plane, and for an in-depth knowledge of the whole-instrument response, which is essential for the development of the data reduction pipeline.

\section{Optical Configuration}
\label{sec:opt_sys_design}
The Strip telescope, originally developed for the CLOVER experiment \cite{North2008}, is a Dragonian side-fed dual-reflector system with a projected diameter aperture of 1.5 m. This design gives exceptionally low aberrations and low cross-polarization across a large focal plane, allowing a large number of detectors to be fed directly. The Dragonian configuration is the best in terms of polarization purity and symmetry over a wide focal region \cite{Dragone1978}.

\begin{figure}
    \centering
    \includegraphics[width=0.8\textwidth]{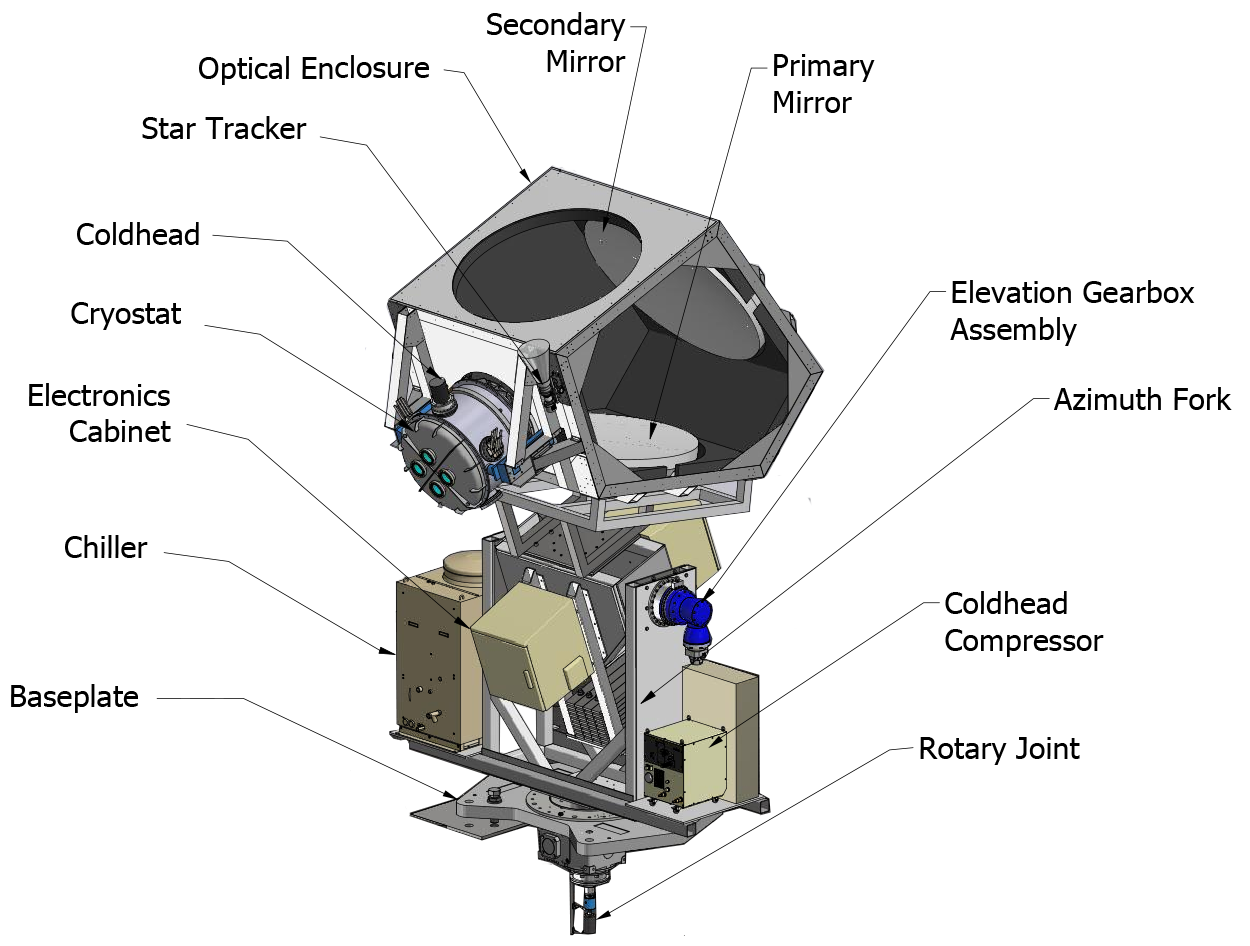}
    \caption[Model of Strip telescope]{Model of the Strip telescope. The mirrors are held inside a co-moving baffle, which is lined with absorber to reduce the effect of sidelobes. In the picture, we omitted one of the baffling panels to allow the view of the mirror arrangement. A counter balance will be mounted on the opposite side of the elevation axis from the telescope. All of the instrument hardware shown here is either built or under construction.}
    \label{fig:telescope_mount}
\end{figure}

The telescope \cite{Villa2021} is surrounded by a co-moving enclosure structure acting as a baffle, made of aluminum plates internally coated with a millimetre-wave absorber, which reduces the cavity effects and multiple reflections, thus minimizing the straylight contamination. The optical assembly, constituted by the two mirrors, the baffle and the detectors, is installed on top of an alt-azimuth mount, which allows the rotation of the telescope around two perpendicular axes, allowing us to change the azimuth and elevation angle. An integrated rotary joint will transmit power and data to the telescope and the instrument, and it will enable a continuous spin around the vertical axis as required by the scanning strategy \cite{Incardona2018}. Figure~\ref{fig:telescope_mount} shows a general view of the Strip telescope.

Hereafter we describe the main optical elements that comprise the optical system and that we used to perform the optical analysis.

\subsection{Feedhorns}
The telescope is illuminated by the array of dual-profiled corrugated feedhorns, where the corrugation profile is a mixture of a sine-squared section, starting from the throat, and an exponential section near the aperture plane. The forty-nine Q-band feedhorns are arranged in a honeycomb lattice of seven hexagonal modules, each including seven feedhorns. The W-band single feedhorns are placed around the Q-band modules in the focal plane of the telescope. The feed array, together with the RF chains, is integrated in a cryostat and cooled down to 20 K. For a detailed description of the Strip feedhorns see \cite{Franceschet2021}.
The performance of the array in terms of return loss has been tested after a cooling cycle, showing no significant variations in the response.

\begin{figure}
    \centering
    \includegraphics[width=0.6\textwidth]{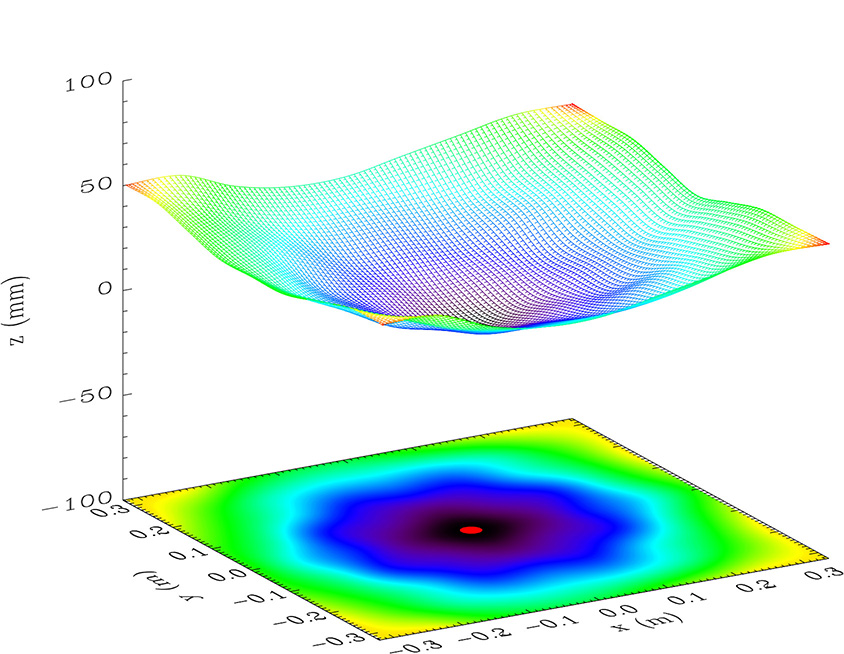}
    \caption[Focal surface of the Strip dual-reflector telescope]{Focal surface of the Strip dual-reflector telescope. The color scale emphasizes the height along the z-coordinate. The reference coordinate system has the origin in the focus of the telescope, which is represented by the red dot.}
    \label{fig:foc_surf}
\end{figure}

The feedhorns are placed on the telescope focal surface (see Fig.~\ref{fig:foc_surf}), which has been computed using the software package WaFER (Wave Front Error evaluatoR) that implements a method to define and characterize the focal surfaces of millimeter wave telescopes by minimizing the variance of the optical path lengths weighted with a feedhorn pattern \citep{WaFER}. 

Each hexagonal feedhorn module is oriented to center the telescope illumination on the primary mirror, so that an optimum spillover is obtained while guaranteeing low level of cross‐polarization contamination. The feedhorn modules are also focused along their axis to optimize the near sidelobe response, i.e. we searched for the position with the lowest near lobes level. The same is true for the W-band feedhorns. The full focal plane array is symmetric with respect to the telescope axis of symmetry.

\begin{figure}
    \centering
    \includegraphics[width=0.55\textwidth]{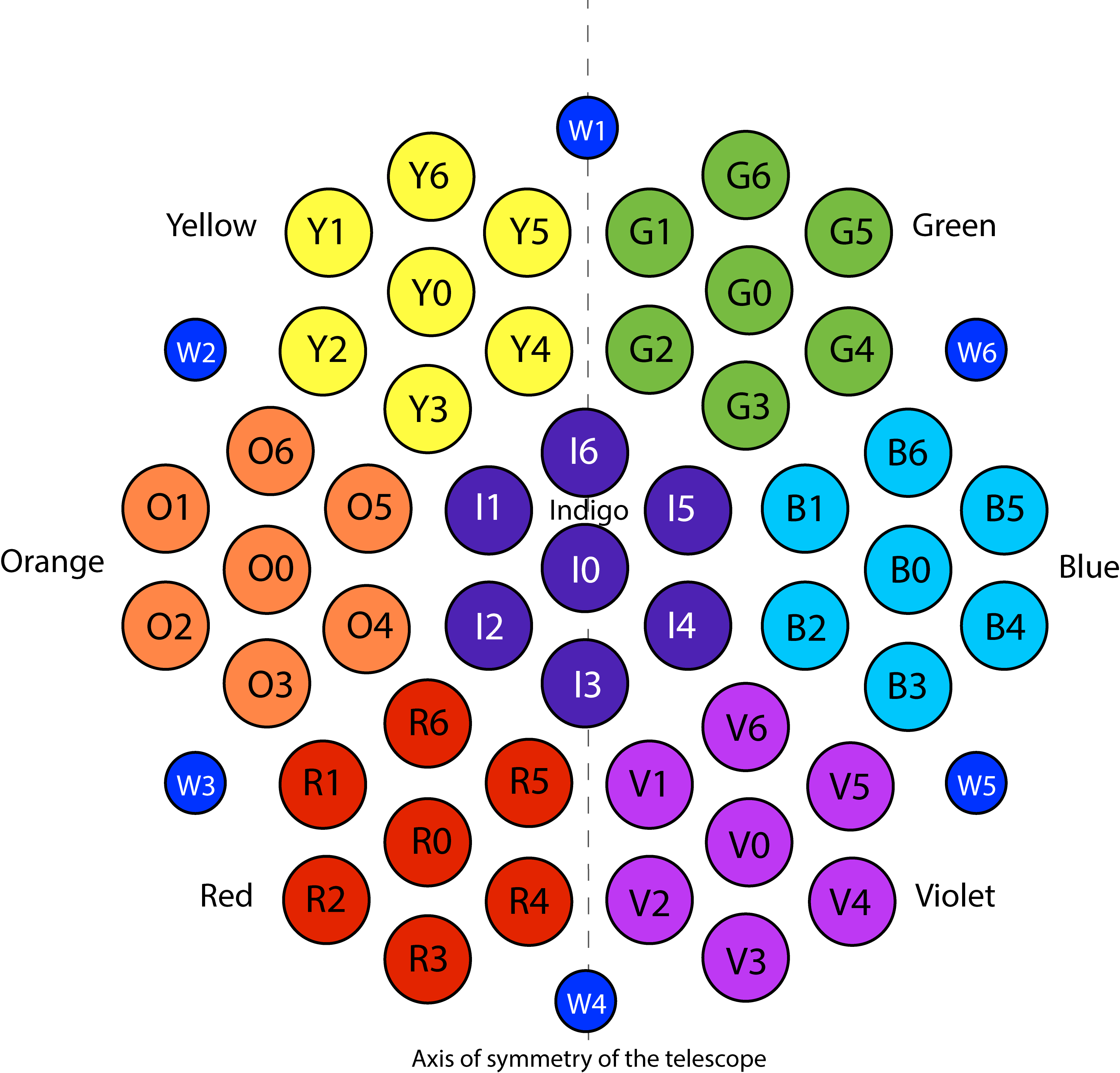}
    \caption[Strip feedhorns naming convention]{Strip feedhorns naming convention. Each Q-band feedhorn is identified by the capital letter of the module color, followed by a number from 0 to 6 in the specified order. Each W-band feedhorn is identified by W followed by a progressive number from 1 to 6. The plane of symmetry of the telescope passes through the vertical axis of the array.}
    \label{fig:naming_fp}
\end{figure}

We decided to use a naming convention in order to identify each feedhorn on the focal plane. We identify the seven modules with a color of the rainbow (Red, Orange, Yellow, Green, Blue, Violet, Indigo), and we associate a number to each feedhorn in a module, so that each horn antenna is identified by the  capital letter of the color followed by a number from 0 to 6; e.g. the central horn of the central module is called I0. The reference focal plane layout is reported in Fig. \ref{fig:naming_fp}. 

We performed simulations of the feedhorns radiation pattern with the software SRSR-D, which provides a rigorous simulation of the electromagnetic performance of any structure with symmetry of revolution, and we used these results to define the source of the optical chain in the GRASP software.

\subsection{Cryostat window and filters}
Infra-red (IR) filters are placed in front of feedhorns to minimize the IR radiation entering the cryostat. They are modelled as perfectly matched dielectric slabs (plane parallel lenses) with their dielectric constant, $\epsilon_r = 2.32$, both in the Q- and W-band, and loss tangent, $\tan \delta  = 10^{-4}$ in the Q-band and  $\tan \delta = 1.3\cdot 10^{-4}$ in the W-band. These numbers are typical of Ultra-High Molecular Weight Polyethylene (UHMWPE) material. The design foresees one filter with a $6$ mm thickness in front of each feedhorn for the W-band horn and one filter $17.32$ mm thick for each of the 7-feedhorns modules in the Q-band, for a total of 13 IR filters inside the cryostat.
A window in front of the filters acts as a vacuum and optical interface between the inner and outer part of the cryostat. This window has been modeled as a slab $43$ mm thick with the same UHMWPE characteristics. 

The filters and the window consist of a core homogeneous slab with a triangular coating on both side, which is used to increase the throughput by reducing reflections at a level better than $-27$~dB, while maintaining very low levels of cross polarization ($<-30$~dB). The thickness used in the simulations is the same of the slab without the anti-reflection coating.

\subsection{Mirrors configuration}

\begin{figure}
    \centering
    \includegraphics[width=1.0\textwidth]{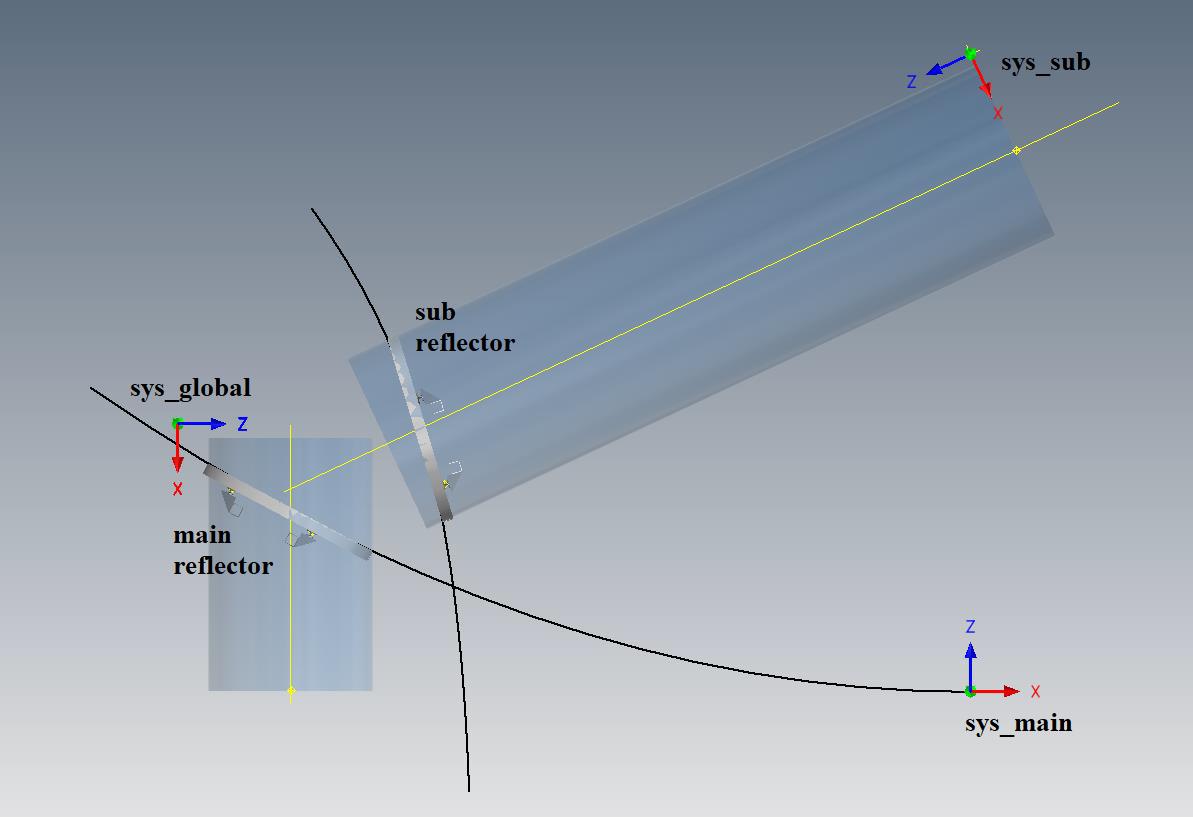}
    \caption[reflector layout]{Layout of the two reflectors of the Strip telescope, together with the originating conics and the relevant coordinate systems used to define the surface of the mirrors. The $sys_\text{main}$ and $sys_\text{sub}$ are the two coordinate system in which the main- and sub-  reflectors are defined; the red arrows are the x-axes, while the blue arrows are the z-axes. The $sys_\text{global}$ is the global reference coordinate system whose origin is placed at the geometrical focal point of the telescope and coincides with the focal plane reference system.}  
    \label{fig:refl_layout}
\end{figure}

The Strip telescope consists of two reflectors arranged in a Dragonian cross-fed design whose layout is shown in Fig.~\ref{fig:refl_layout}. The relative position of the coordinate systems needed for the description of the layout is reported in Table \ref{tab:sys_position}.
\begin{table}[]
\centering
\begin{tabular}{lccc}
\hline
&$x$ (mm)& $y$ (mm)&$z$ (mm)\\
\hline
$\text{sys}_\text{main}$ & 2430.3536 &0.0 & 7205.146935\\
$\text{sys}_\text{sub}$ &-3359.81512& 0.0 &7205.146935\\
\hline
\end{tabular}
\caption{\label{tab:sys_position}Origin of the main and sub-reflectors coordinate systems given in the global frame.}
\end{table}

The main reflector surface is defined as paraboloid of revolution in the $sys_\text{main}$ coordinate system. The originating parabola has a focal length of $f_\text{main}=5790.17$ mm and the surface is defined by the following equation: 

\begin{equation}
z_\text{main} = \frac{x_\text{main}^2 + y_\text{main}^2 }{{4\cdot f_\text{main}}}\text{.}
\end{equation}

The main reflector rim is defined as the intersection between the mirror surface and a cylinder in the $z_\text{main}$ axis direction with a cross section given in the $(x,y)_\text{main}$-plane. The projection of the rim onto this plane is circular with the center placed at $xc_\text{main}=-6180.34$~mm and $yc_\text{main}=0.00$~mm and with an half-axis of $750$~mm, which provides a full $1.5$~m aperture and thus an angular resolution of $\sim 20'$ in the Q-band and $\sim 10'$ in the W-band. 

The sub-reflector is defined as an hyperboloid of revolution in the $sys_\text{sub}$ coordinate system, and it is defined by the equation
\begin{equation}
z_\text{sub} = c - \frac{a}{b} \sqrt{\left ( b^2 + x_\text{sub}^2 + y_\text{sub}^2 \right )}\text{,}
\end{equation}
where the vertex distance is $2a = 3845.33$ mm, and the foci distance is $2c=7950.00$ mm, which result in $b=3479.08$ mm. 
The sub-reflector rim is defined as the intersection between the mirror surface and a cylinder in the $z_\text{sub}$ axis direction with a cross section given in the $(x,y)_\text{sub}$-plane. In this plane, the rim is elliptical with the half-axis of $859.88$~mm and $829.35$~mm along $x_\text{sub}$ and $y_\text{sub}$, respectively. The center is placed at $xc_\text{sub}=967.43$~mm and $yc_\text{sub}=0.00$~mm in the $sys_\text{sub}$ coordinate system.

The angle between the paraboloid and hyperboloid axes is $25^{\circ}$, allowing the focal plane to point in the horizontal direction when the line of sight of the telescope points in the vertical direction. 

The entire dual-reflector system has an equivalent focal length of 2700~mm, resulting in an F-number 1.8. This is a typical value for Dragone-Mizuguchi crossed configurations because of the specific design that imposes a long equivalent focal length to have the secondary focus accessible.

The shielding structure surrounding the reflector is made of aluminum plates coated by a millimetre-wave absorber. It has been modelled as a series of plane reflectors with the appropriate rim, except for the top panel, which is characterized by the presence of a circular aperture. The aperture has a radius of $770$ mm and it has been modelled as an aperture stop in an infinite conducting screen. Thus, the scattering from the aperture is determined by means of Babinet’s principle \citep{Ticra:2008}. We did not consider the absorber, so the model represents the worst case.

\begin{figure}
    \centering
    \includegraphics[width=0.7\textwidth]{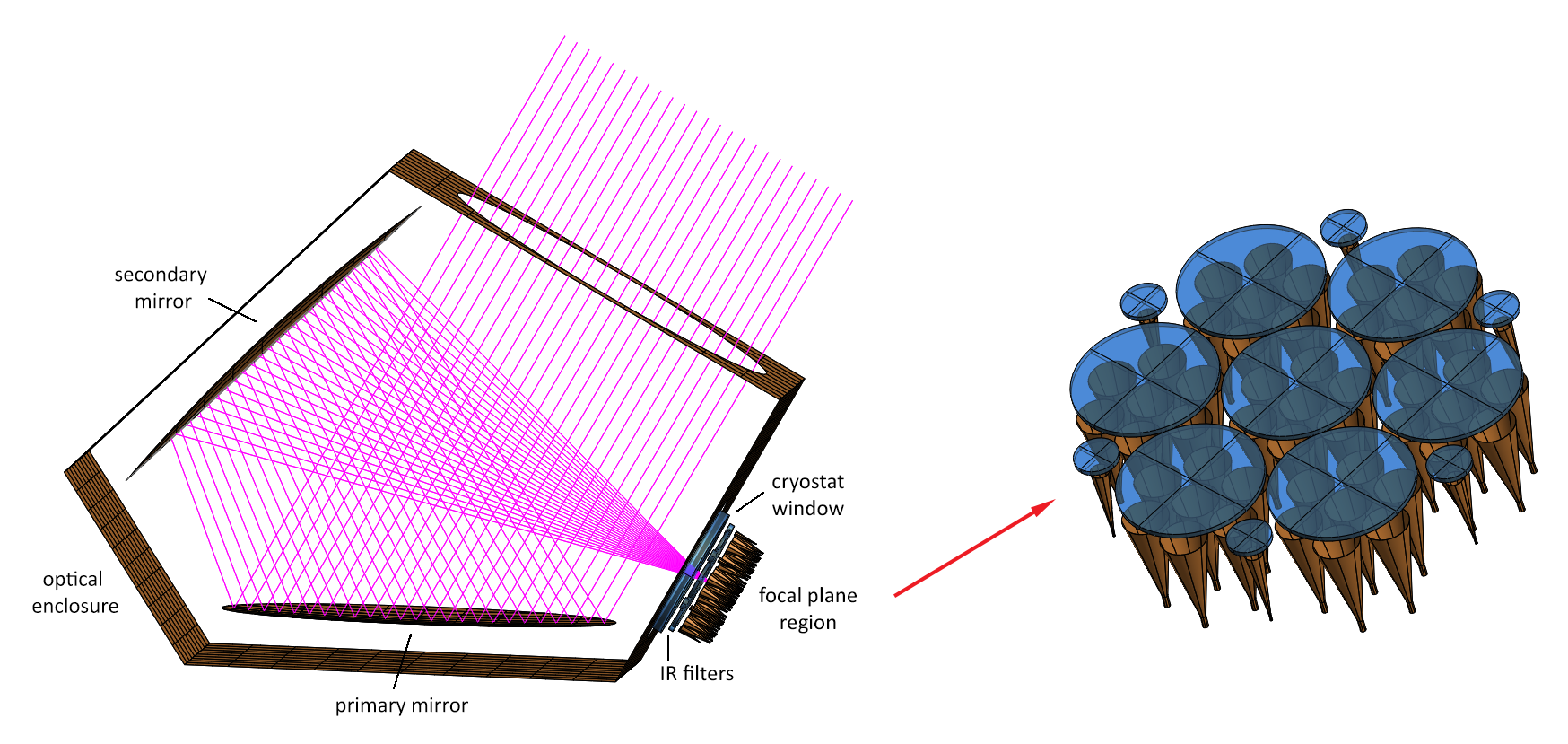}
    \caption[Electromagnetic model of the Strip telescope]{Electromagnetic model of the Strip telescope. It includes the parabolic main reflector (MR), the hyperbolic sub-reflector (SR), the forty-nine Q-band, the six W-band feedhorns, the shielding structure with the circular aperture, the cryostat vacuum window and the IR filters. The rays show the ideal optical path of radiation when the central feedhorn is considered as the source.}
    \label{fig:tel_model}
\end{figure}

As shown in Fig.~\ref{fig:tel_model}, the incoming radiation is reflected from the primary mirror, then from the secondary mirror and finally reaches the feedhorns placed in the focal plane of the optical system.

\section{Optical performance}
The Strip optical system has been modelled with the software GRASP (version 10.6.0), developed by TICRA, which is the standard design tool for reflector antennas. We performed optical simulations considering the feedhorn as a source and computing the pattern scattered by both reflectors in the far-field.

\subsection{Simulation method}
Nowadays, several electromagnetic simulation methods are available, like Physical Optics (PO), Physical Theory of Diffraction (PTD), Geometrical Optics (GO), and Geometrical Theory of Diffraction (GTD). These methods are widely used to perform robust electromagnetic simulations and can be used for beam prediction of real optical systems.

We performed main beam simulations using Physical Optics and Physical Theory of Diffraction on both reflectors. Physical Optics is a simple method that gives an approximation to the surface currents and is valid for perfectly conducting scatterers which are large in terms of wavelengths.

On the one hand, the PO method is the most accurate to predict beams propagating from  reflector antenna systems and it may be used in all angular regions of the space surrounding the system. On the other hand, as the frequency increases the reflectors have to be more precisely sampled, and simulations time can increases impracticably.

For this reason, we computed the sidelobes using the Multi-Reflector Geometrical Theory of Diffraction (MrGTD) as implemented in GRASP, which computes the scattered field from the reflectors performing a backward ray-tracing \citep{Nielsen:2000}. This method has been extensively used to simulate the sidelobes of the ESA Planck mission \citep{sandri2010}. 
The purpose of the MrGTD is to calculate the GTD fields from any number of reflectors illuminated sequentially starting at a given source. Once we have defined the reflector geometry, the source characteristics, and the output field points, together with each contribution (i.e., a bundle of rays defined by a sequence of scatterers and by the type of interaction -- reflection or diffraction -- on each of them) to be taken into account, we can reach an accurate radiation pattern prediction.

For each path connecting the source to the far-field direction, we test the fulfilment of the diffraction or reflection laws. After we find the ray traces, we look for possible shadows from all the defined structures. If no such intersections exist, we compute the field by the standard GTD method for each reflection or diffraction points.
When many scatterers are involved, the amount of ray tracing contributions may lead to unacceptable computational time even with MrGTD. Since each contribution to the sidelobes is computed separately, it is crucial to identify the contributions which produce significant power levels in the resulting radiation pattern.

We tested the time required for the simulation of the sidelobes of a single feedhorn considering a different number of interactions (reflections or diffractions). We found that it takes approximately a week for a simulation up to the second order, whereas, moving to the third order can require one month. However, since the power level associated with third order of interaction is negligible, we decided to simulate only up to the second order.

\subsection{Main beams}
We computed far-field radiation patterns in the co- and cross-polarization basis according to Ludwig's third definition \cite{Ludwig1973} in $uv$-spherical grids to retrieve the major electromagnetic characteristics.
The $uv$-grid defines field points in a 2D grid on a sphere where the field will be calculated. The 2D grid is defined by the ($u$, $v$) coordinates, where $u$ and $v$ are dimensionless and can be related to the spherical angles $\theta$ and $\phi$ of
the direction to the far-field by
\begin{equation}
\begin{gathered}
    u = \sin \theta \cos \phi \text{,}\\
    v = \sin \theta \sin \phi \text{.}
\end{gathered}    
\end{equation}

\begin{figure}
    \centering
    \includegraphics[width=0.49\textwidth]{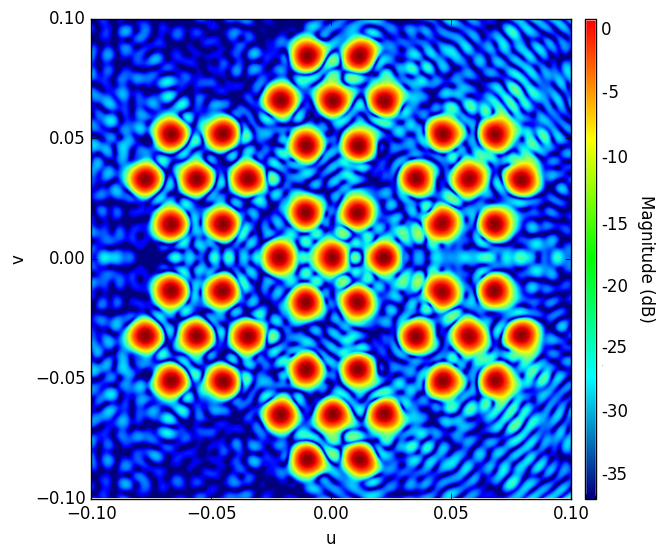}
    \includegraphics[width=0.49\textwidth]{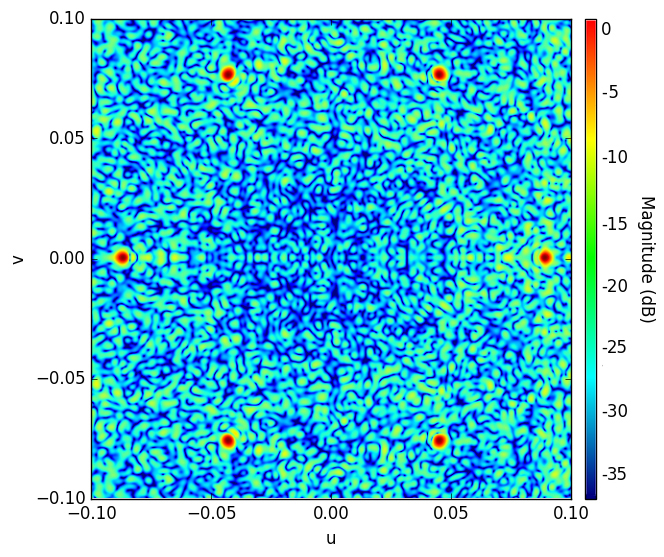}
    \caption[Footprint of beams in the sky]{\emph{Left}: Footprint of the forty-nine Q-band beams in the sky in ($u$, $v$) coordinates. \emph{Right}: Footprint of the six W-band beams in the sky. The telescope symmetry plane is along the $v$-axis. The spots in the upper part of the left plot are related to the R, O and Y modules (from left to right). These footprints show that the overall width of the main beams in the sky is $\sim 10^{\circ}$.}
    \label{fig:footprint}
\end{figure}

The feedhorns are excited individually with a linearly polarized signal at the simulation frequency, so that the forty-nine Q-band and six W-band radiation patterns can be calculated. The footprint of the 49 Q-band beams and 6 W-band beams in the sky is shown in Fig.~\ref{fig:footprint}. The overall width of the footprint is $\sim 10^{\circ}$.

The main beams for all the forty-nine Q-band feedhorns have been simulated within the range $-0.02<(u, v) < 0.02$, corresponding to a range between $\pm 1.15^{\circ}$ in $\theta$ and $\phi$ polar coordinates. Each grid is sampled with $601 \times 601$ points, which means a spatial resolution of about 14 arcsec.

Given the telescope configuration and the feedhorn off-axis location on the focal surface, the main beams are distorted and their shape differs from the Gaussian symmetric one. For this reason, the main beams cannot be mathematically represented by a single parameter (e.g. the Full-Width Half-Maximum of a Gaussian curve). For a complete characterization of the main beams, we evaluated several descriptive parameters: the angular resolution (FWHM), the ellipticity, the main beam directivity, the cross-polarization discrimination factor (XPD). The FWHM is calculated taking the arithmetic average value between the maximum and minimum of the FWHM of the distorted beam and the ellipticity is defined as the ratio of these values. The XPD, usually expressed in dB, has been computed as the ratio between the maximum directivity of the co– and cross–polar components.

Table~\ref{tab:mb_param} summarizes the major characteristics of the main beams, while a more detailed description of each beam is reported in Appendix \ref{app:mb_par}. The 
The beams associated to the off-axis receivers are more subject to beam distortions; however, all beams are compliant with the requirements on the optics polarization purity, showing XPD $>30$~dB. 
\begin{table}[htbp]
\centering
\begin{tabular}{lcc}
\hline
&43 GHz& 95 GHz\\
\hline
Angular resolution & 20.7$^\prime$--21.3$^\prime$ & 9.3$^\prime$-- 9.5$^\prime$\\
Directivity & 54.4--54.7 dBi & 61.4--61.6 dBi\\
XPD & 40.8--44.5 dB & 44.1--46.6 dB\\
Ellipticity & 1.001--1.028 & 1.006--1.041\\
\hline
\end{tabular}
\caption{\label{tab:mb_param} Typical main beam descriptive parameters at the central frequency of the Q- and W-bands.}
\end{table}

\paragraph{Edge taper evaluation}
The contour plots of the total amplitude field incident on the main reflector have been computed for each Strip Q-band feedhorn, in a surface grid with $301 \times 301$ points. The total amplitude field is defined as $\sqrt{E_x^2+E_y^2+E_z^2}$, where ($x$, $y$, $z$) is the coordinate system in which the field is calculated. The $z$-axis is pointing to the main beam direction.

\begin{figure}
    \centering
    \includegraphics[width=0.95\textwidth]{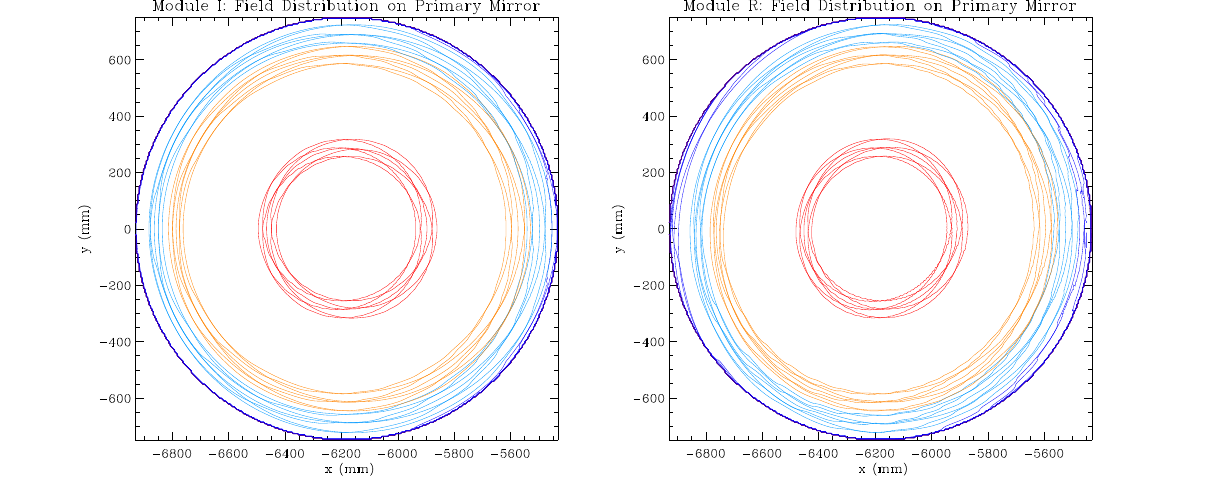}
    \caption[Field distribution on the primary mirror]{Field distribution on the primary mirror for the module in the focus of the telescope (left) and an external one (right). Contour levels at $-3$ dB (red), $-15$ dB (orange), $-20$ dB (light blue), and $-30$ dB (blue) are plotted.}
    \label{fig:et_main}
\end{figure}

\begin{figure}
    \centering
    \includegraphics[width=0.9\textwidth]{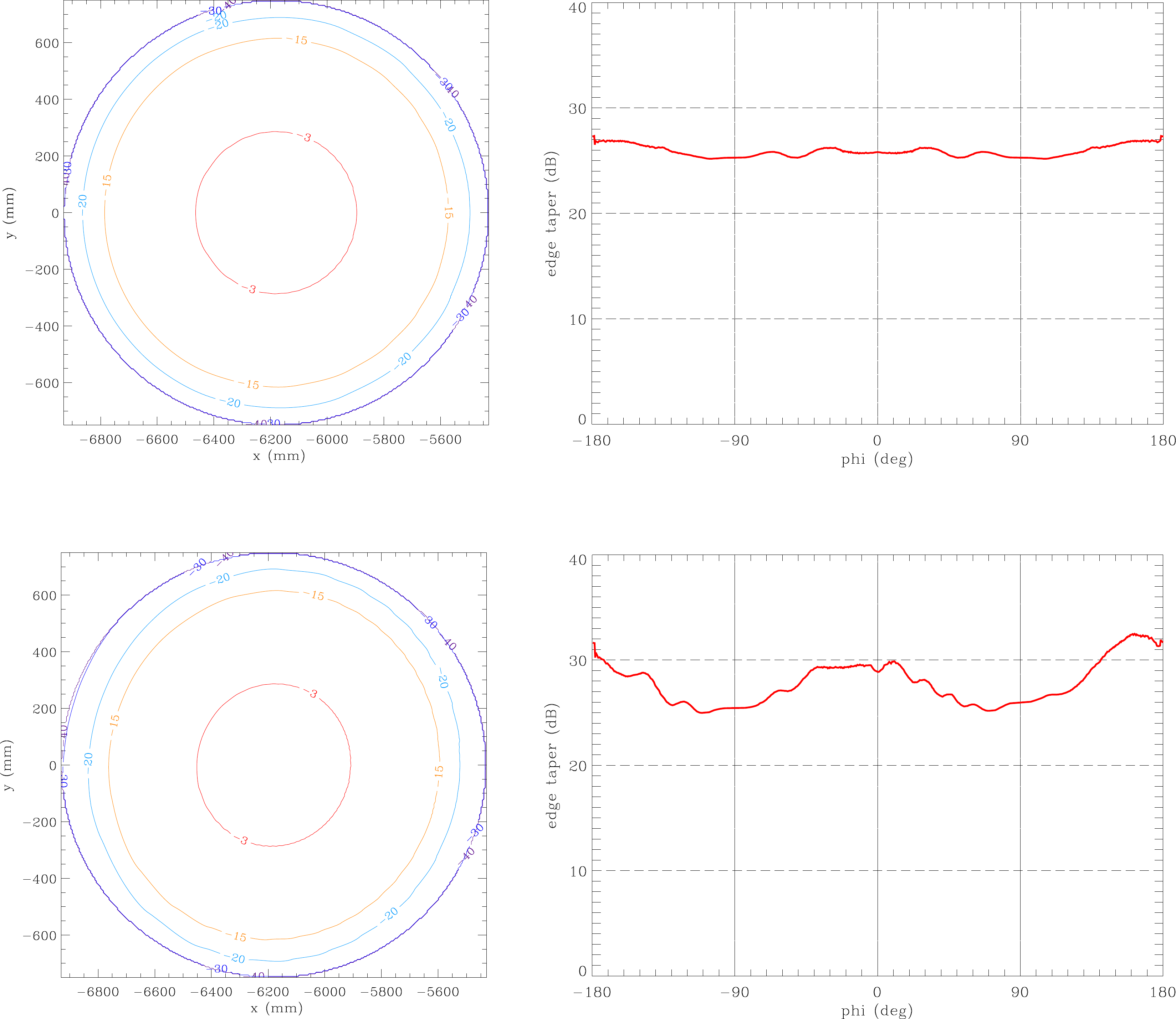}
    \caption[Edge taper curves for I0 and R0]{Field distribution on the primary mirror (left) and edge taper curves (right) for the central feedhorn of the module I (top row) and for the central feedhorn of the module R (bottom row) at 43 GHz. Contour levels at $-3$, $-15$, $-20$ and $-30$ dB (red, orange, light blue, and blue respectively) are plotted.}
    \label{fig:et_IR}
\end{figure}

Figure~\ref{fig:et_main} shows the field distribution on the primary mirror for the module in the focus of the telescope and an external one. From the figure we can see that the feedhorns have been correctly placed so that the illumination is centered on the primary mirror. However, the illumination is roughly elliptical; hence, the field amplitude on the primary mirror rim is not constant. Figure~\ref{fig:et_IR} shows the edge taper curves for the central feedhorns of the same modules reported previously. The angle $\phi$ moves along the rim of the mirror counterclockwise.

\paragraph{Effects of telescope imperfections}
The theoretical shape of the reflectors can be approached only up to some finite limit tolerance set by fabrication constrains. Moreover, deformations can occur during telescope operation due to the effect of gravity, temperature variations and wind pressure.

Irregularities on the reflector surfaces induce phase variations across the antenna aperture and the resulting performance of the optical system could be degraded. For these reasons, we evaluated the impact of such distortions on the main beam. 

In particular, we evaluated the effect of mechanical tolerance as a random error on the surface, whereas we analysed the effect of a structured variation due to external effects describing the surface by means of Zernike polynomials. The analysis of the telescope structural behavior provided the Zernike polynomials expansion used for the subsequent optical analysis.

\subparagraph{Random variations.}The surface of the mirrors can be defined in a regular grid with random $z$-values, specified by a correlation distance and an amplitude. 
The surface values at the nodes of the regular grid are selected as random numbers uniformly distributed in an interval defined by the peak value and with a mean value equal to zero. At intermediate points, the surface is determined by a cubic interpolation. The rms-value of the generated surface, due to the cubic interpolation algorithm, is given by $0.47\cdot\text{peak}$. Since the Strip mirrors are designed to have a peak error within $50\,\mu$m, we get an rms-value of 23.5 $\mu$m.

The spacing between the nodes relative to the reflector diameter determines the roughness of the surface. We considered three different scales for the random distortions analysis: the scale of the central wavelength ($7.46$ mm), a variation on the scale of the whole telescope and an intermediate scale of $\sim 20$ cm, which should take into account variations due to the mirrors supporting structure. Each realization has the same rms-value of 23.5 $\mu$m. The percentage variation of main beam descriptive parameters computed with ideal reflectors and with random distortions for four representative horns shows that variations are lower than 0.5\%. 

\subparagraph{Structured variations.}We can define the surfaces through a weighted sum of Zernike polynomials. With this description of deformations, we can analyse with GRASP the effect of the structural behavior of the main and sub-reflector of the Strip telescope for different design loads and configurations. The structural analysis of the reflectors has been made considering the mirrors as supported on a rigid structure; hence, it takes into account only the reflector stiffness. The present study does not consider the influence from the optical enclosure, the telescope mount and the baseplate, which will be evaluated in a forthcoming analysis.
The elevation range of the telescope, and corresponding mirrors orientation, is between $35^{\circ}$ and $90^{\circ}$ given as elevation angles (i.e. between $55^{\circ}$ and $0^{\circ}$ zenith angles)\footnote{Elevation angles are measured from the horizontal position of the telescope, while zenith angles are measured from the zenith.}. 

Different loads act or may act on mirrors during telescope operative conditions. We considered gravity, rotation and movements effects, thermal loads (temperature variation and gradient) and wind pressure.
Electromagnetic simulations show that the percentage variation of main beams FWHM computed with ideal reflectors and with distortions is lower than 0.05\%. The same holds true for ellipticity, with the exception of the load of strong winds (speed greater than 12.5 m/sec\footnote{The typical wind speed expected at the observation site will range between 2 and 10 m/s}), which show a greater distortion with respect to all other loads. We noted also that the effect of gravity is stronger for higher elevation angles. Furthermore, a shift in the pointing direction due to mirror distortions may occur and this effect, together with other structural effects and telescope misalignment, will be carefully analysed by the LSPE-Strip team.

\subsection{Sidelobes}
Straylight contamination may be one of the most critical sources of systematic effects during observations. For this reason, we analysed the $4\pi$ beams of the Strip feedhorns coupled with the dual-reflector telescope, taking into account its shielding structure. A detailed understanding of the level of the sidelobes and of their origin is an essential input for the data analysis. 

\begin{figure}
    \centering
    \includegraphics[width=0.9\textwidth]{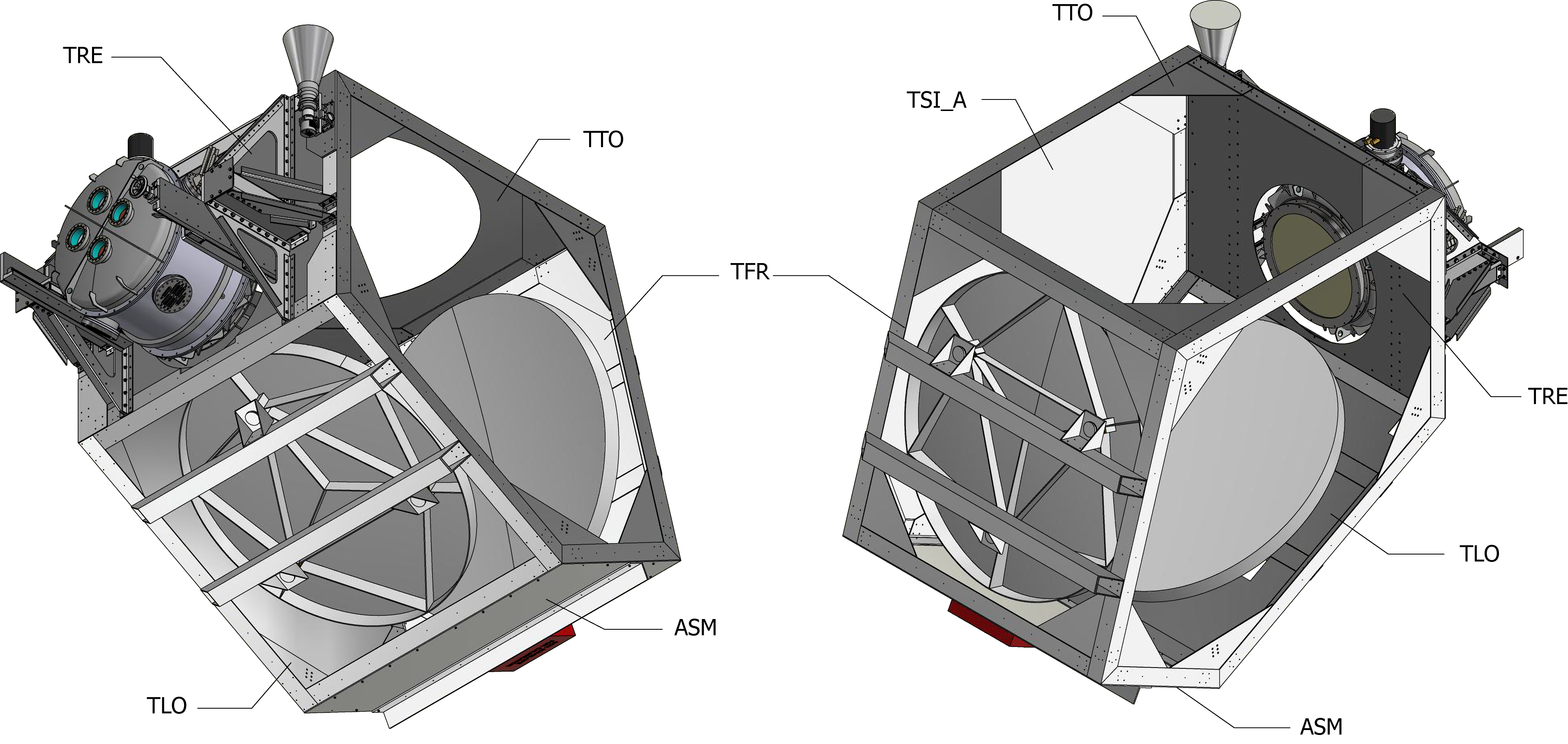}
    \caption[Mechanical design of the shielding structure]{Mechanical design of the shielding structure, together with the nomenclature adopted to describe each panel. In the pictures, we omitted some of the baffling panels to allow the view of the mirror arrangement.}
    \label{fig:montatura}
\end{figure}

The Strip telescope is surrounded by a co-moving baffle made of seven aluminum plates of different shapes in order to reduce the contamination due to the sidelobes. 
Proper evaluation of the effect of reflective shields is crucial, since these structures redistribute the power that is radiated by the feedhorns and is not reflected by the telescope.
Figure \ref{fig:montatura} shows the mechanical design of the shielding structure and the nomenclature adopted to describe each panel.

In our simulations, the two reflectors and the shielding structure have been considered as blocking elements. The shielding panels have been modelled as perfectly conducting surfaces and we did not consider the triangular corner plates which allow the panels to be mounted on the telescope structure to reduce the complexity of the model.
Using the MrGTD approach, each contribution to the sidelobes is computed separately, accounting for interactions at different orders. Each interaction can be a reflection or a diffraction.
We computed the sidelobes of the Strip telescope up to the 2\textsuperscript{nd} order of interaction at 43 GHz for four representative channels. We computed a total of 240 contributions for each channel.

\begin{figure}
    \centering
    \includegraphics[width=\textwidth]{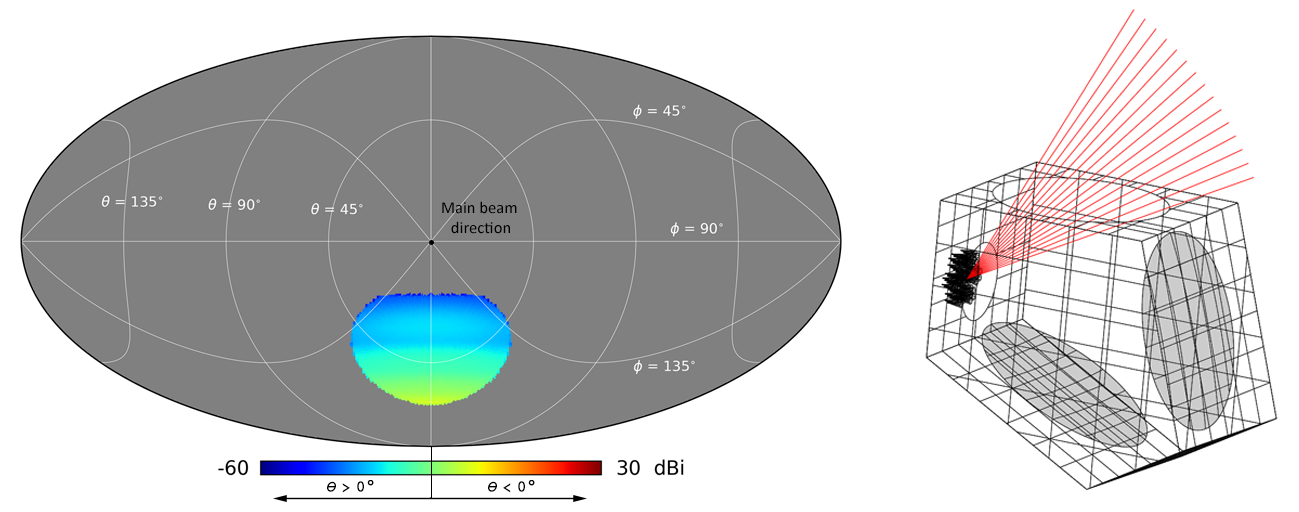}
    \caption[Scheme of the direct contribution to the sidelobes]{\emph{Left}: $4\pi$ map of the field due to the rays coming from the feedhorn (direct contribution) is shown. Most of the map is empty (gray colour) because most of the rays are blocked by the baffle. The power peak of the direct contribution is -62.65 dB below the main beam directivity. \emph{Right}: Sketch of the optics with the ray-tracing of the direct contribution.}
    \label{fig:direct_rays}
\end{figure}

\begin{figure}
    \centering
    \includegraphics[width=\textwidth]{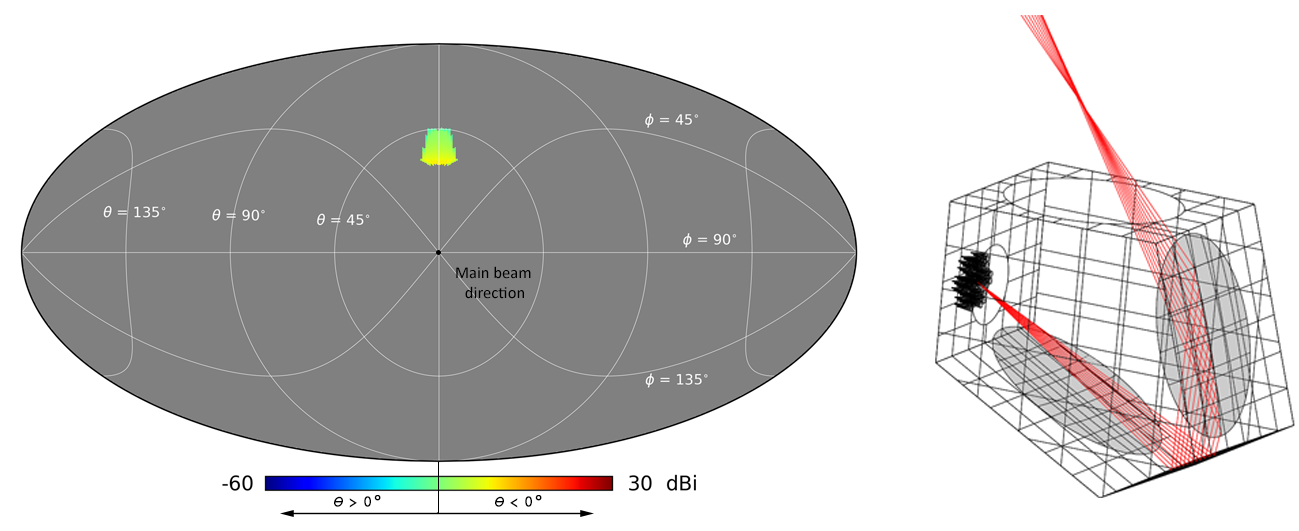}
    \caption[Scheme of the ASMrSr contribution to the sidelobes]{\emph{Left}: $4\pi$ map of the field due to the rays that are reflected on the ASM panel and then reflected by the sub-reflector (namely ASMrSr contribution). The power peak of the direct contribution is about -58.46 dB below the main beam directivity. \emph{Right}: Sketch of the optics with the ray-tracing of 2\textsuperscript{nd} order contribution.}
    \label{fig:2refl_rays}
\end{figure}

We sample the field with a $0.5^{\circ}$ step both in the $\theta$ and $\phi$ coordinates. Sidelobes are represented as full-sky Mollweide projections using the HEALPix \cite{Healpix1999} visualization facilities, where the main beam axis points towards the North pole of the maps. As an example, Fig.~\ref{fig:direct_rays} shows the co-polar contribution to the sidelobes due to radiation coming directly from the feedhorn and not intercepted by the reflecting structures for the feedhorn in the telescope focus at 43 GHz. Figure~\ref{fig:2refl_rays} shows a 2\textsuperscript{nd} order contribution consisting of a series of two reflections, the first on a shielding panel and the second on the sub-reflector. 

\begin{figure}
    \centering
    \includegraphics[width=0.9\textwidth]{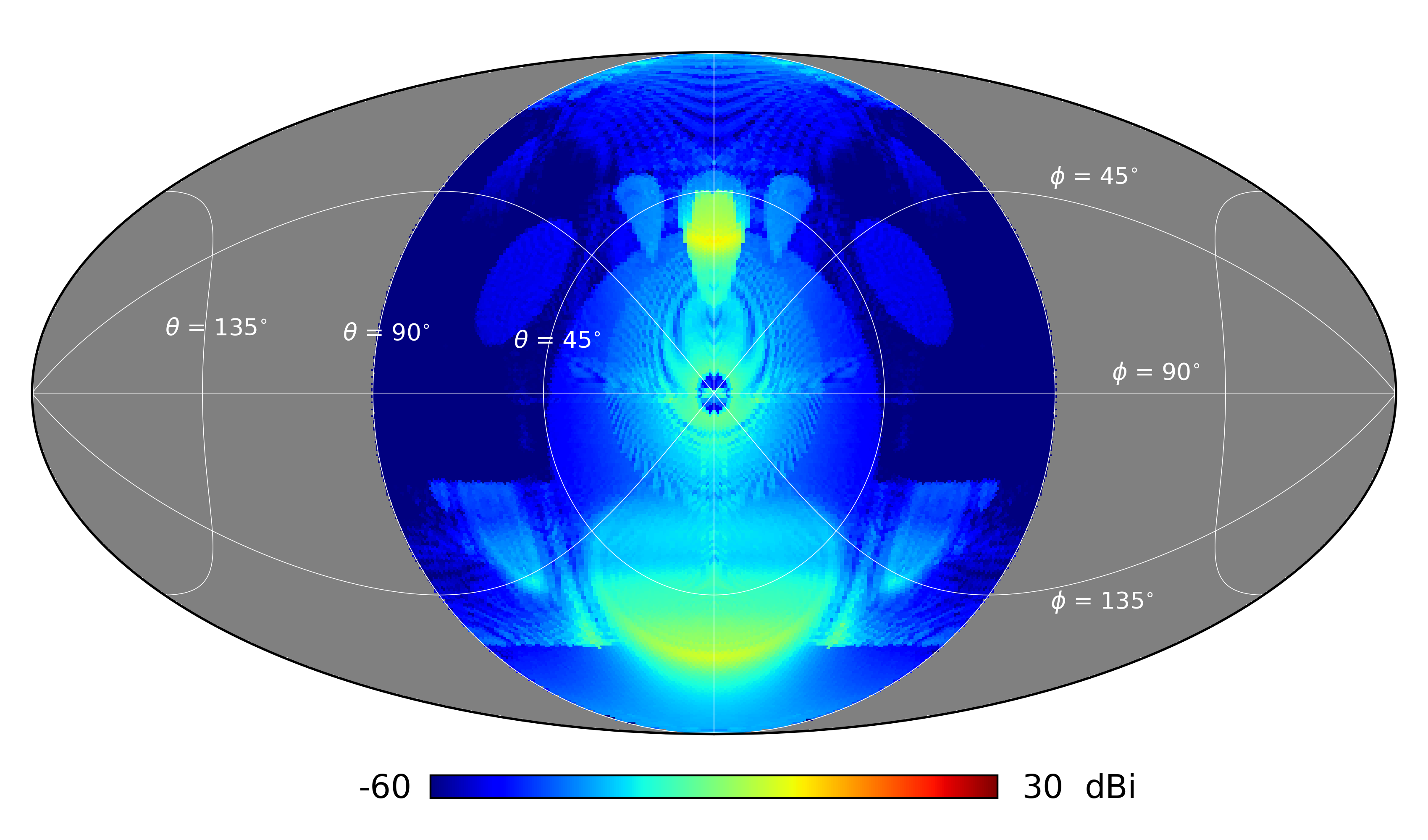}\\
    \includegraphics[width=0.9\textwidth]{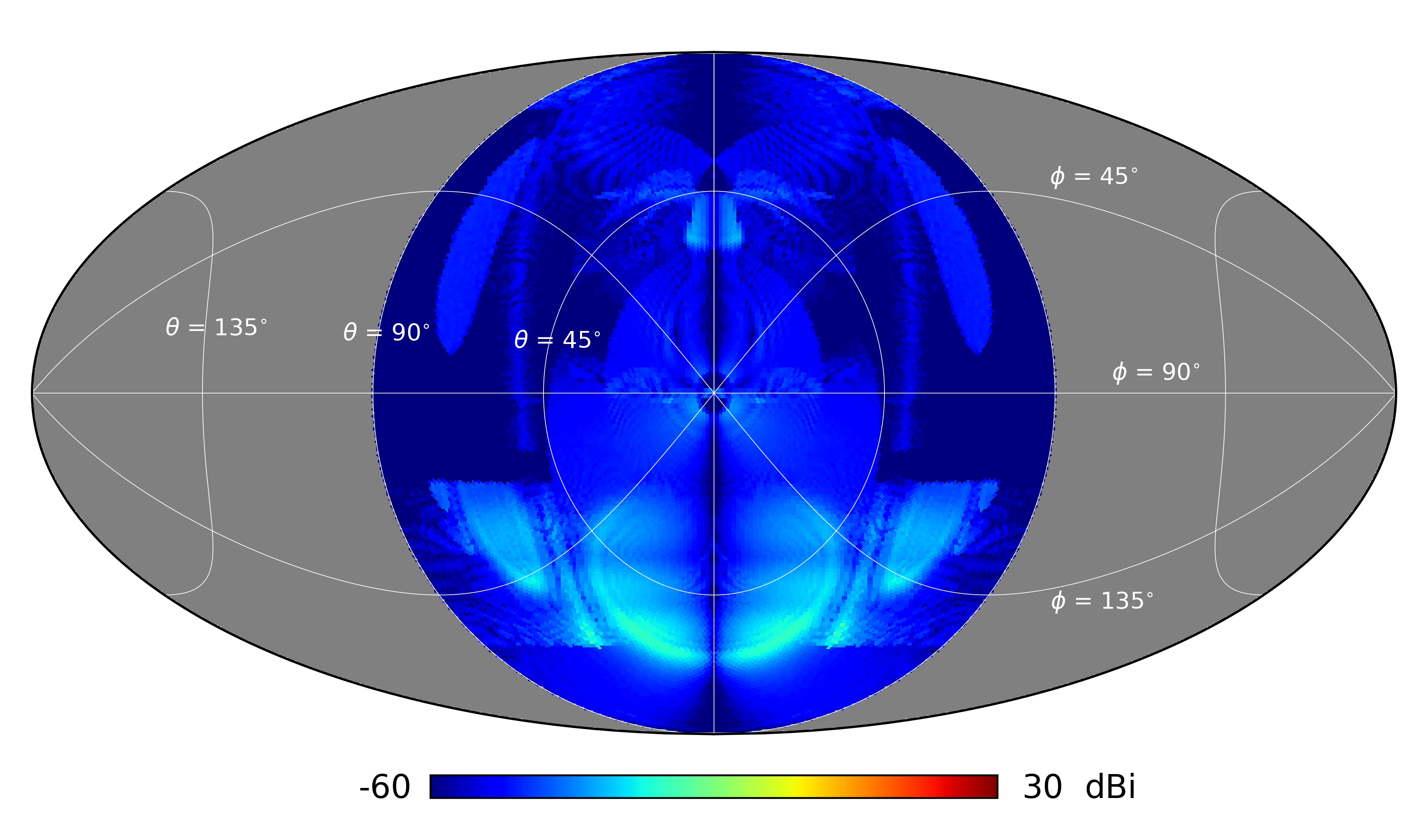}
    \caption[I\textsubscript{0} co- and cross-polar radiation pattern over the full sky]{Co–polar (\emph{top}) and cross-polar (\emph{bottom}) component of the radiation pattern for the central feedhorn computed with MrGTD at 43 GHz over the full sky.}
    \label{fig:sl_I0}
\end{figure}

\begin{figure}
    \centering
    \includegraphics[width=0.9\textwidth]{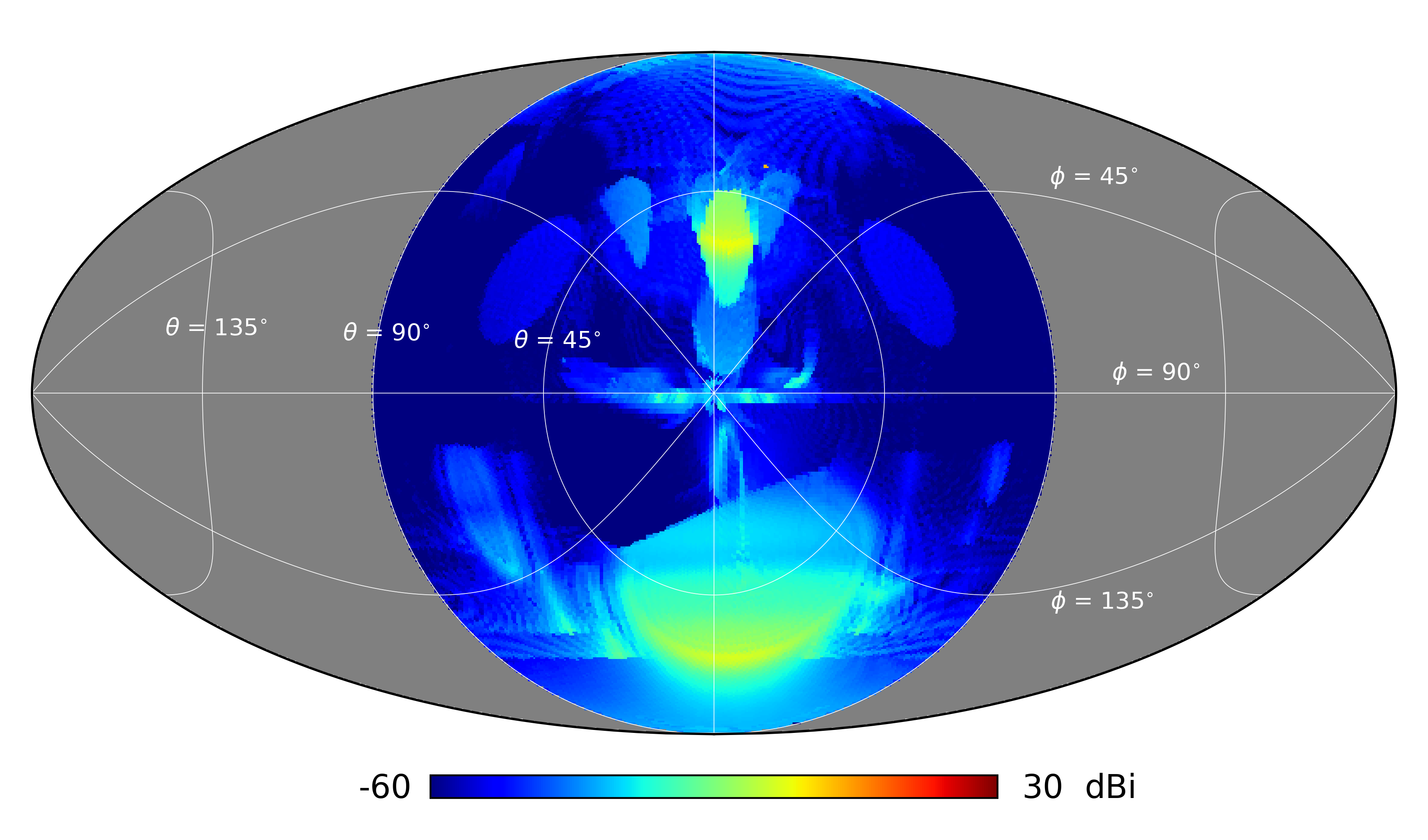}\\
    \includegraphics[width=0.9\textwidth]{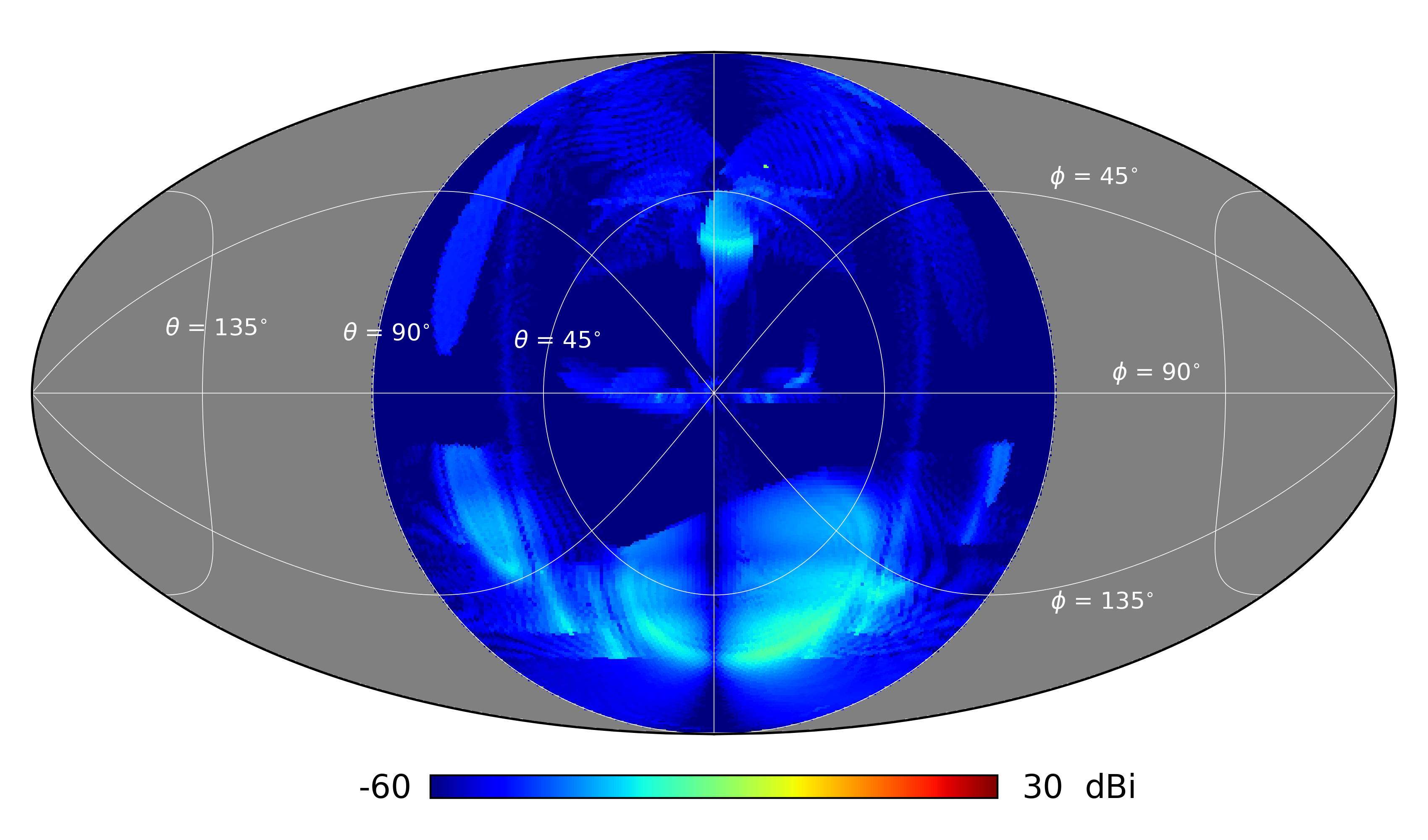}
    \caption[O\textsubscript{2} co- and cross-polar radiation pattern over the full sky]{Co–polar (\emph{top}) and cross-polar (\emph{bottom}) component of the radiation pattern for an external feedhorn (O2) computed with MrGTD at 43 GHz over the full sky.}
    \label{fig:sl_O2}
\end{figure}

After calculating every single contribution at different orders, we summed them up to get the complete sidelobes map for each channel.
Figures~\ref{fig:sl_I0} and \ref{fig:sl_O2} show the final co-polar and cross-polar maps for the central 43 GHz feedhorn (I0) and an external one (O2). The maximum power level in the sidelobes with respect to the main beam directivity is -58.69 dB and -53.17 dB, respectively. The lower half of the full-sky map is empty because of the presence of a panel with a circular aperture on top of the shielding structure; since here we use a ray-tracing technique, rays directions are limited within $\theta = \pm 90^{\circ}$.

The Mollweide projections show that the sidelobes are unevenly distributed and concentrated mainly in two areas, namely the direct contribution and the double reflection on the ASM panel. The direct contribution is generated by the rays entering the feedhorns without any interaction with the reflectors; its shape and power level are given by the feedhorn radiation pattern pointing at about $60^{\circ}$ from the telescope reference boresight. On the other hand, the double reflection is primarily due to rays reflected by one of the shields (ASM panel) and then reflected by the sub-reflector. These are the brightest contributions in the sidelobe region. 
The power level of the two contributions is comparable in the four feedhorn analysed. We can clearly see that the double reflection affects an angular region narrower than the direct contribution, but it is closer to the main beam. 

The knowledge and possibly the reduction of the sidelobe level is still of major importance. In fact, the telescope sidelobes will pickup the 300~K emission from the ground, which could be polarized. For this reason, we are planning to cover the inside structure with an absorbing material to minimize the sidelobes contribution. Since a complete analysis that accounts also for the absorbers is very challenging, it was not addressed here and it will be performed only after the full analysis of the sidelobes impact on Strip observations. 

\section{Conclusions}
Beams characteristics have a strong impact and play a key role on observations: unwanted optical aberrations such as main beam asymmetries due to the off-axis configuration of the telescope, spurious effects of the cross--polarized component, as well as the not--negligible level of sidelobes, inevitably impact on the measurements of the CMB polarization signal.

A comprehensive knowledge of the beam radiation pattern would in principle allow the reconstruction of the true signal from the sky, within the accuracy limits imposed by systematic effects. For this reason, we developed a detailed electromagnetic model of the Strip optical system and we performed simulations to determine its optical response.

Our model includes the feedhorns, the two mirrors, the shielding structure with a circular aperture, and the cryostat infrared filters and window. We characterized the response of the optics both in the main beam and sidelobe region. 

We found a very uniform behaviour of the main beams, with dispersion in FWHM lower than 0.3$^\prime$ and excellent symmetry both in Q- and W-bands. 
We studied also the effect of mirrors imperfections, modelled as random errors with a given rms and structured deviation described by means of Zernike polynomials expansion. We found that the effect of these imperfections on main beams parameters is lower than 1\%, and thus they are negligible. 

We analysed also the sidelobes of the telescope taking into account the presence of the comoving baffling structure, which significantly modifies the radiation pattern shape far from the telescope boresight.
Using the Multi-Reflector Geometrical Theory of Diffraction method, we identified the sequences of reflections and diffractions which give a significant contribution to the sidelobes. In particular, we found that two contributions lead to a power level in the sidelobes that should be carefully considered in data analysis and systematic effects evaluation.

The results of these simulations will represent an invaluable tool for assessing the performance of the Strip telescope at system level.

\acknowledgments
The LSPE--Strip instrument has been developed thanks to the support of ASI contract I/022/11/1 and Agreement 2018-21-HH.0 and by funding from INFN (Italy). We would like to thank also the company BCV progetti s.r.l., which performed the analysis of the telescope structural behavior.

\appendix
\section{Strip main beam parameters}
\label{app:mb_par}
The following tables summarize the major characteristics of the main beams of each feedhorn in half the focal plane due to the telescope symmetry. The angular resolution is calculated as the arithmetic average value between the maximum and minimum of the resolution of the  distorted beam. We computed also the geometric mean, but there is no significant difference with the arithmetic average reported in the table. The ellipticity is defined as the ratio of the maximum and minimum values of the resolution. The XPD is the ratio between the maximum directivity of the co– and cross–polar components.
\begin{table}[h]
    \centering
    \begin{tabular}{ccccccc}
    \toprule
Beam	&	FWHM	($'$)	&	e	&	$\mathcal{D}$	(dBi)	&	XPD	(dB)	\\
\midrule												
I0	&	20.70	&	1.006	&	54.71	&	44.48	\\			
I1	&	20.74	&	1.002	&	54.69	&	43.28	\\			
I2	&	20.76	&	1.009	&	54.68	&	44.44	\\			
I3	&	20.78	&	1.013	&	54.67	&	43.93	\\			
I4	&	20.76	&	1.009	&	54.68	&	44.44	\\			
I5	&	20.74	&	1.002	&	54.69	&	43.28	\\			
I6	&	20.72	&	1.005	&	54.70	&	43.19	\\			
\midrule												
Y0	&	20.88	&	1.003	&	54.58	&	41.55	\\			
Y1	&	21.06	&	1.006	&	54.48	&	40.80	\\			
Y2	&	20.94	&	1.005	&	54.56	&	42.29	\\			
Y3	&	20.81	&	1.005	&	54.63	&	42.59	\\			
Y4	&	20.80	&	1.006	&	54.63	&	42.86	\\			
Y5	&	20.92	&	1.000	&	54.56	&	41.43	\\			
Y6	&	21.06	&	1.004	&	54.48	&	41.01	\\			
\midrule												
O0	&	21.01	&	1.014	&	54.51	&	44.29	\\			
O1	&	21.18	&	1.017	&	54.42	&	43.99	\\			
O2	&	21.20	&	1.019	&	54.41	&	44.08	\\			
O3	&	21.08	&	1.016	&	54.48	&	43.69	\\			
O4	&	20.91	&	1.009	&	54.57	&	43.81	\\			
O5	&	20.88	&	1.010	&	54.58	&	43.49	\\			
O6	&	21.01	&	1.011	&	54.51	&	43.84	\\			
\midrule												
R0	&	21.08	&	1.025	&	54.50	&	42.38	\\			
R1	&	21.10	&	1.022	&	54.49	&	43.37	\\			
R2	&	21.28	&	1.027	&	54.39	&	41.84	\\			
R3	&	21.30	&	1.028	&	54.39	&	41.49	\\			
R4	&	21.13	&	1.024	&	54.47	&	41.55	\\			
R5	&	20.96	&	1.022	&	54.57	&	43.31	\\			
R6	&	20.93	&	1.022	&	54.58	&	43.55	\\			
\bottomrule
    \end{tabular}
    \caption[Main beam characteristics at the central frequency 43 GHz]{Main beam characteristics at the central frequency 43 GHz.}
    \label{tab:mb_center_Q}
\end{table}

\begin{table}[h]
    \centering
    \begin{tabular}{ccccc}
    \toprule
Beam	&	FWHM\textsubscript{A} ($'$)	&	e	&	$\mathcal{D}$ (dBi)	&	XPD (dB)	\\
\midrule
W1	&	9.41	&	1.041	&	61.45	&	44.43	\\
W2	&	9.34	&	1.027	&	61.52	&	46.57	\\
W3	&	9.32	&	1.011	&	61.56	&	46.53	\\
W4	&	9.49	&	1.006	&	61.40	&	44.14	\\
    \bottomrule
    \end{tabular}
    \caption[Main beam characteristics at the central frequency 95 GHz]{Main beam characteristics at the central frequency 95 GHz.}
    \label{tab:mb_center_W}
\end{table}


\bibliographystyle{ieeetr} 
\bibliography{biblio} 

\end{document}